# Characterizations of individual mouse red blood cells parasitized by *Babesia microti* using 3-D holographic microscopy

HyunJoo Park[1,+], Sung-Hee Hong[2,+], Kyoohyun Kim[1,+], Shin-Hyeong Cho[2], Won-Ja Lee[2], Youngchan Kim[1], Sang-Eun Lee[2,*] & YongKeun Park[1,*]

[1]Department of Physics, Korea Advanced Institute of Science and Technology, Daejeon 305-701, Republic of Korea.

[2]Division of Malaria & Parasitic Diseases, Korea National Institute of Health, Korea Centers for Disease Control & Prevention, Chungbuk 363-951, Republic of Korea.

[+]These authors contributed equally to this work.

[*]To whom correspondence should be addressed. E-mail: yk.park@kaist.ac.kr (Y.P.) or ondalgl@korea.kr (S.-E.L.)

## Abstract

Babesia microti causes "emergency" human babesiosis. However, little is known about the alterations in *B. microti* invaded red blood cells (*Bm*-RBCs) at the individual cell level. Through quantitative phase imaging techniques based on laser interferometry, we present the simultaneous measurements of structural, chemical, and mechanical modifications in individual mouse *Bm*-RBCs. 3-D refractive index maps of individual RBCs and in situ parasite vacuoles are imaged, from which total contents and concentration of dry mass are also precisely quantified. In addition, we examine the dynamic membrane fluctuation of *Bm*-RBCs, which provide information on cell membrane deformability.

Babesiosis, a tick-borne zoonosis from rodents and mainly caused by intraerythrocytic *Babesia* protozoa, has been an important disease of livestock and recently emerged as a health threat to human throughout the world. *Babesia* spp. are major significances in economy and health issues, and provide unique opportunities for a better understanding of the pathophysiology of malaria due to their similarities including pathologic symptoms, intraerythrocytic cycle, erythrocyte cytoadherence, and cytokine production[1-3]. *B. microti* is an obligate parasite which causes "emergency" babesiosis in human, hamster, and mouse, which invades mature red blood cells (RBCs)[4,5], and the pathophysiology of *B. microti* is strongly related to the intraerythrocytic stage, which causes significant alterations in host RBCs[6]. However, our understanding of the alterations in *B. microti*-invaded RBCs (*Bm*-RBCs) are mostly unexplored and still remains incomplete, which is largely due to the limitations of experimental techniques.

Currently, imaging techniques for the study of *Bm*-RBCs is generally based on optical microscope with Giemsa-stained blood smears[7,8] and the discrimination of parasites, including their morphological and pathological features still relies on the experiences of professional technicians. However, these conventional approaches only provide limited structural information and Giemsa staining only allows the visualization of the nucleic acids of parasites; whole parasitized vacuoles (PVs) are not visualized. Alternatively, electron microscopic techniques has been used[9], but it requires a difficult cell fixation process and a metal coating, and thus the live dynamics of cells and parasites cannot be determined. Furthermore, fixation artifacts including loss of parasite membranes have been reported[10].

These limitations of conventional imaging techniques complicate assessments of the characterization of *Bm*-RBCs, especially at the individual cell level. For the past few years, however, there has been an emergence of quantitative phase-imaging (QPI) techniques, based on optical interferometry or digital holography, which can offer quantitative, non-invasive, and label-free imaging of biological cells and tissues[11,12]. In particular, the recently developed common-path diffraction optical tomography (cDOT) can measure both the 3-D structure and dynamics of live cells simultaneously[13].

In this article, employing cDOT in mouse infection model, the alterations in mouse *Bm*-RBCs are quantitatively and precisely characterized at a single cell level. In order systematically to investigate *Bm*-RBCs, measurements of the morphological, biochemical and mechanical parameters through the quantitative imaging of individual

RBCs and *in-situ* parasites are performed in three groups: healthy RBCs, *Bm*-RBCs having apparent PVs, and *Bm*-RBCs without apparent PVs. 3-D refractive index (RI) distributions of *Bm*-RBCs reveal the morphological modifications in host RBCs with the occupation of PVs. The biochemical alterations in *Bm*-RBCs, such as contents and concentration of the cellular dry mass, are also quantified. Furthermore, dynamic membrane fluctuations, which are strongly correlated with the mechanical deformability of cell membrane, are simultaneously measured at the individual cell level. We also show that invading *B. microti* parasites use the cytosolic ATP of host cells for their metabolism. The present method provides unique advantageous for the study of *Bm*-RBCs and facilitate a better understanding of the pathophysiology of babesiosis.

**RESULTS**

**3-D RI Maps of *Bm-RBCs***

To obtain the 3-D optical tomograms of individual *Bm*-RBCs, we quantitatively measured the 3-D RI distributions of *Bm*-RBCs, employing cDOT[13,14]. cDOT is a QPI technique which, as noted above, provides simultaneous measurements of 3-D RI maps and dynamics in the cell membrane of individual cells. First, we measured the 3D distribution of RI, $n(x, y, z)$, of RBCs from three healthy mice and three infected mice with *B. microti* having 20–30% parasitemia, determined by Giemsa staining (15 days after infection). RI, an intrinsic optical property of a material, provides quantitative imaging contrast in QPI. For the parasitemia used in this study, infected RBCs show mainly the ring form of the infection, while the shape and number of PVs inside individual RBCs vary.

As shown in Fig. 1, 3-D RI maps of *Bm*-RBCs clearly reveal the shapes and locations of the PVs as well as overall shapes of host RBCs, whereas healthy RBCs exhibit characteristic donut shapes or discocytes. These PVs exhibit RI values lower than that of RBC cytosol, indicating that *B. microti* creates its own environment inside the cytosol of a RBC and depletes the hemoglobin (Hb) in PVs. 3-D RI maps of *Bm*-RBCs are measured in various infection stages depending on the number of PVs inside a host cell: single (Fig. 1B), double (Fig. 1C), quadruple (Fig. 1D), multiple (> 3) PVs (Fig. 1E) and without PVs (Fig. 1F), and a healthy RBC (Fig. 1A). The arrows in Figs. 1B-E indicate the locations of individual or clustered PVs. Intriguingly, in the *B. microti*-infected RBC, we observed no pigment hemozoin, a disposal product formed from the metabolism of Hb by some blood-feeding parasites such as malaria *Plasmodium falciparum*[15,16]. The absence of hemozoin is one of characteristics of *Babesia* spp. that differentiates it from *P. falciparum*[17].

**Structures of *in situ* PVs**

The detailed structures of *in situ* PVs inside host cells are examined in the measured 3-D RI maps. In general, spherical PVs are found in RBCs having a single PV or PVs apart from others (Figs. 1B-C). As the number and size of the PVs increase, however, PVs often show branched shapes or a digitiform (Fig. 1G). Furthermore, the 3-D RI maps allows the quantitative structural analyses of PVs, including the precise locations inside the cells, the sizes, the volume, and the RI, whereas conventional diagnostic methods such as Giemsa staining or exogenous fluorescent labeling only provide limited morphological and nucleic acids information about parasites[4,17]. The rendered isosurfaces of the measured RI tomograms ($n = 1.350$) effectively visualize the overall size, shapes and locations of PVs. It is clearly shown that individual PVs in a host cell have different sizes and locations (e.g., the top and side views of the isosurfaces in Fig. 1).

The measured morphologies and sizes of PVs are consistent with those measured with confocal microscopy or blood smear[4,17]. The sizes of individual PVs range is 1–3 fL, which is only 2–5% of infected RBC volume (approx. 60 fL). For infected RBCs with multiple PVs (e.g. Fig. 1E), the total volume of PVs occupies as much as 30% of a whole cell. The averaged RI of individual PVs is in the range of 1.348–1.375, which is lower than that of the cytoplasm of healthy RBCs ($n_{RBC} = 1.390$) and higher than that of a buffer medium ($n_m = 1.334$). The increased RI of the PVs indicates an increase in the molecular concentration, implying the production of parasite proteins. Separate PVs in the same host cell exhibit different RI values. For example, the averaged RI for PVs located on the bottom and right side of the cell in Fig. 1C1 are 1.360 and 1.375, respectively. This indicates asynchronous asexual budding of a trophozoit into merozoites[5].

In addition, the structure of PVs forming a cross was also examined. Inside the host RBCs, merozoites become trophozoites and undergo binary fission; connected parasites in 'paired forms' are hallmark of babesiosis[7]. As shown in Figs. 1G-J (the same cell in Fig. 1D), the cross-sections of RI tomograms along different planes clearly reveal three PVs adjacent to one another, which may be physically connected at the upper location of the host cell because each parts of PVs exhibit similar level of RI. On the *x-y* plane in focus, the three PVs are also located closely to each other (Fig.1G). At 1 μm above the focal plane, however, the PVs seem to be merged into one object, whereas they are separated at 1 μm below the focal plane (Figs. 1H-J). However, it should be noted that due to the optical diffraction limit (the lateral and axial resolution of cDOT was experimentally calibrated as 270 and 920 nm, respectively), vacuoles located closer than the diffraction limit are hardly resolved and it may appear as one large cluster from the top view (Figs.1 D2 & E2). In addition, thin structures such as the center dimple area of healthy RBCs or thin gap between a host cell membrane and a PV are barely imaged and appear as holes in the rendered RI isosurface.

**Structural alterations of host RBCs due to *B. microti* invasion**

To investigate the structural alterations in *Bm*-RBCs, we quantitatively analyzed the 3-D RI maps and retrieved important red cell indices, including cellular volume, surface area, and sphericity index, as shown in Fig. 2. The invasion of *B. microti* resulted in an increase in the cellular volume. The mean values of the cellular volume are $48.2 \pm 8.9$, $58.9 \pm 14.2$, and $57.2 \pm 14.2$ fL for healthy RBCs, *Bm*-RBCs with occupied PVs and those without PVs, respectively (Fig. 2A). These results obtained with cDOT are consistent with independently measured MCV (mean corpuscular volume, $45.6 \pm 6.3$ and $49.0 \pm 8.9$ fL for healthy RBCs and infected RBCs, respectively) using a complete blood count (CBC) blood test (gray dot lines in Fig. 2A), which measures cell volumes using electrical impedance. It is important to note that values measured with CBC test are averaged from whole blood, whereas those measured with cDOT are from individual RBCs. Compared to the CBC test, the mean value of the cellular volume deviates slightly from that measured by cDOT but both measurements show that the cellular volume consistently increases as a result of an invasion by *B. microti*.

In contrast, the surface areas of RBCs do not exhibit significantly changes upon parasite invasion, while the standard deviation of the surface areas increases slightly in *Bm*-RBCs both in the presence and absence of *B. microti* PVs (Fig. 2B). The mean values of the surface areas are $102 \pm 13$, $107 \pm 21$, and $106 \pm 22$ μm$^2$.

After an invasions of *B. microti*, the RBC sphericity indices, a dimensionless parameter which describes the sphere-likeness of a 3-D object (1 for a perfect sphere and 0 for a flat surface), exhibit significant increases (from 0.6 to 0.7). This indicates that *Bm*-RBCs undergo morphological alterations from discocytes (healthy RBCs) to spherocytes or sphere-like RBCs by losing their dimple regions, which is similar to when ATP is derived inside RBCs[18,19]. These morphological changes are observed mostly in the presence of PVs.

**Changes in membrane deformability of host RBCs due to *B. microti* invasion**

To investigate the alterations of mechanical properties of RBCs due to the parasitization by *B. microti*, the dynamic membrane fluctuations of *Bm*-RBCs were quantitatively measured. From the measured 2-D dynamic optical phase-images of individual RBCs, $\Delta\phi(x, y, t)$, the dynamic cell height information, $h(x, y, t)$, can be retrieved considering the RI of individual RBCs obtained from the 3-D RI maps as $h(x, y, t) = [\lambda/(2\pi\cdot\Delta n)]\Delta\phi(x, y, t)$, where $\lambda$ is the wavelength of a laser, and $\Delta n = \langle n(x, y, z)\rangle - n_m$ is a RI difference between cell cytoplasm $\langle n(x, y, z)\rangle$ and surrounding medium $n_m$. The dynamic membrane fluctuation of RBCs measured using cDOT is more precise than that obtained with other existing QPI techniques such as diffraction phase microscopy[20,21], because the RI of RBC cytoplasm can be taken into account at the single-cell level in cDOT, whereas previous techniques assumes the same $\Delta n$ for RBCs. The topographical images, calculated as the mean cell heights, $h_m(x, y) = \langle h(x, y, t)\rangle$, of healthy RBCs, *Bm*-RBCs with PVs and *Bm*-RBCs without PVs are presented in Figs. 3A-C. For areas occupied with PVs, the cell heights appear to be lower, resulting from the fact that the RI of the PVs are lower than that of the cytoplasm. For *Bm*-RBCs, cells with PVs exhibit a loss of the center dimple regions, whereas cells without PVs maintain their characteristic donut shapes.

To address the mechanical deformability of the cell membrane of *Bm*-RBCs, the dynamic membrane fluctuations were quantitatively analyzed. From the measured cell height information, the root-mean-square (*RMS*) of height displacement (see Methods) was obtained at each point on the cell. The *RMS* height displacement can be translated as the deformability or elasticity of the cell membrane cortex[13,22,23]. The *RMS* height fluctuations of representative cells for the healthy RBCs, *Bm*-RBCs with PVs, and *Bm*-RBSs without PVs are presented in Figs. 3D-F, respectively. The *RMS* height displacements of the *Bm*-RBCs are compatible with those of healthy RBCs, indicating that the deformability of *Bm*-RBCs does not significantly change due to the invasion of

*B. microti* parasites. Translational motions of individual PVs inside host RBCs are also observed for *Bm*-RBCs with PVs, as depicted with high values of the *RMS* height displacements with ring shapes (as indicated by the arrows in Fig. 3E).

To further analyze the deformability of the membrane of *Bm*-RBCs, the membrane fluctuations were determined from averaged *RMS* height displacements over cell areas (Fig. 3G). For a direct comparison of the deformability of the cell membrane only, the area occupied by PVs in *Bm*-RBCs were excluded for the analysis. It is clearly seen that there is no statistical difference in the membrane deformability between healthy RBCs and *Bm*-RBCs. The mean values of the membrane fluctuations are 57.1 ± 9.9, 59.5 ± 11.9, and 59.0 ± 13.6 nm.

**Cellular dry mass and mass concentration in *Bm*-RBCs**

To quantitatively investigate the cellular dry mass, i.e., the non-aqueous materials inside cell [24] of *Bm*-RBCs, we used the quantitative phase images obtained by cDOT. Because the concentration of the cellular dry mass is linearly proportional to the RI via a parameter known as the RI increment, the total cellular dry mass (CM) of *Bm*-RBCs including PVs are obtained by integrating the 2-D optical-phase images of individual cells over the cell area and applying the values of RI incensement[24-26]. An RI increment of 0.2 ml/g for Hb was used for the analysis [27]; it is known that there is no significant difference in the values of the RI increase between protein species[28,29]. The mean values of CM are 16.4 ± 3.5, 20.7 ± 6.1 and 21.0 ± 6.9 pg for healthy RBCs, *Bm*-RBCs with PVs, and *Bm*-RBCs without PVs, respectively (Fig. 4A). These results obtained with cDOT are consistent with independently measured MCH (mean corpuscular hemoglobin, 16 and 18.5 pg for healthy RBCs and infected RBCs, respectively) using a CBC blood test (gray dot lines in Fig. 3A). MCH was measured by spectrophotometry using a cyanmethemoglobin procedure, which involves the oxidization of hemoglobin to methemoglobin using excess ferric iron[30]. Compared to healthy RBCs, *Bm*-RBCs exhibit significant increases of CM regardless of the presence of PVs inside. In comparison with the healthy RBCs, the CM increased by 26% and 28% for *Bm*-RBCs with PVs, and *Bm*-RBCs without PVs, respectively, which strongly suggest the production of parasite proteins, since RBCs without a nucleus do not produce proteins by themselves .

The mass concentration of *Bm*-RBCs are also calculated by dividing obtained CM by the cellular volume retrieved from the measured 3-D RI maps at the individual cell level. These results are shown in Fig. 4B. The mean values of the cellular mass concentration are 34.0 ± 4.0, 35.3 ± 6.0 and 36.4 ± 6.6 g/dL for healthy RBCs, *Bm*-RBCs with PVs, and *Bm*-RBCs without PVs, respectively. These results are consistent with MCHC (mean corpuscular hemoglobin concentration, 35.1 and 36.2 g/dL for healthy RBCs and infected RBCs, respectively) using a CBC blood test (gray dot lines in Fig. 3A). MCHC was determined by MCH divided by MCV[30]. Compared to the healthy RBCs, the mass concentrations are not statistically different for the *Bm*-RBCs. Interestingly, the total mass concentrations for *Bm*-RBCs without PVs did significantly increased.

**Discussions**

**Optical micro-tomography for the study of *Bm*-RBCs**

A novel optical micro-tomographic image technique based on laser interferometry is presented for the study of *Bm*-RBCs. This label-free and non-invasive optical technique provides quantitative imaging capabilities, which enable precise and systematic investigations of alterations occurring in individual host cells and *in situ* in the parasite upon a *B. microti* invasion. The present method using cDOT enables simultaneous measurements of structural, chemical, and mechanical alterations at the single cell level, which provides unique advantages in investigating *Bm*-RBCs. It provides precise structural information of *Bm*-RBCs. From measured 3-D RI tomograms, the shapes, volumes, surface areas, and sphericity of the host RBCs are obtained at the single-cell level. The volume, structure, and location of the PVs are also investigated. Chemical information is also quantitatively addressed. The total contents and concentration of the dry mass are obtained from the RI information of *Bm*-RBCs. Furthermore, the mechanical properties of the cell membrane deformability are also be obtained by measuring the dynamic membrane fluctuations in the *Bm*-RBCs, without using conventional contact approaches such as micropipette aspiration, atomic force microscopy, or optical tweezers[31].

**Structural and biochemical alterations in *Bm*-RBCs**

*Bm*-RBCs exhibit substantial alternation in their morphology. Due to the formation of PVs, the total cellular volumes of the *Bm*-RBCs increased by 22%. However, the cell surface areas of *Bm*-RBCs do not significantly increase, implying that *B. microti* does not generate any additional lipid membrane while host RBCs do not undergo significant microvesiculization. Thus, *Bm*-RBCs become more sphere-like from donut-like healthy RBCs, as manifested by an increase of their sphericity indices. The center dimple region is also lost in *Bm*-RBCs, which cannot be solely explained by the increase in the internal volume due to the formation of a PV. It was observed that RBCs maintain the center dimple regions when the internal volume is simply increased by water influx due to osmotic pressure[32]. This suggests that more significant alternations occur at cell membrane cortex. One possibility is ATP depletion; the loss of the dimple shape is one of characteristic hallmarks of ATP depletion in RBC cytoplasm[18,33].

The increases of the total dry mass in *Bm*-RBCs including PVs are prominent. Compared to healthy RBCs, *Bm*-RBCs gain mass at a rate of 26%, which indicates active the metabolic processes by *B. microti* parasites (Fig. 4A). Interestingly, mass concentration of *Bm*-RBCs does not significantly differ from that of healthy RBCs (Fig. 4B). The correlation analysis reveals that the contents and concentration of the cellular dry mass are strongly regulated for *Bm*-RBCs and for healthy RBCs (Fig. 4C). The mass concentrations remain within a range of 34 – 36 g/dL for both healthy RBCs and *Bm*-RBCs.

**Bystander effects of a *B. microti* parasitization**

*Bm*-RBCs without PVs exhibit significant alterations in their structural and chemical properties. In this study, *Bm*-RBCs without PVs include both RBCs that are not invaded by *B. microti* parasites and RBCs that are invaded by *B. microti* parasites, with the parasites being too small to be resolved in cDOT[17,34]. However, it is clear that *Bm*-RBCs without PVs exhibit homogenous distribution of RI in their cytoplasm and that there are is no noticeable structure inside the host RBCs. Nonetheless, the cellular volumes of *Bm*-RBCs without PVs are significantly larger than those of healthy RBCs, and are compatible with the *Bm*-RBCs with PVs (Fig. 2A). The sphericity indices of *Bm*-RBCs without PVs are also compatible with those of *Bm*-RBCs with PVs, and significantly higher than those of healthy RBCs (Fig. 2C). Yet, the center dimple regions remain in *Bm*-RBCs without PVs, unlike *Bm*-RBCs with PVs. This is intriguing because *Bm*-RBCs without PVs lack the inclusion of large PVs whose volume occupies up to 30% of the whole cell volumes in *Bm*-RBCs with PVs. This bystander effect during a *B. microti* infection implies significant pathophysiological alterations in the level of blood circulatory system or higher, not at the individual cellular level.

**Glucose or ATP as a major energy source for *B. microti*?**

Our results indicate that during intraerythrocytic process the metabolism of *B. microti* parasites do not rely on the consumption of Hb protein, unlike malaria-inducing *Plasmodium* spp. The mass concentrations in the cytoplasm of *Bm*-RBCs do not show significant decreases regardless of the intraerythrocytic development stage of the parasites. In addition, the formation of hemozoins are not observed in *Bm*-RBCs, even at the terminal intraerythrocytic stage. However, *B. microti* parasites utilize active metabolic processes considering the significant increases of the cellular dry mass and the formation of large PVs. This suggests that energy resource other than Hb protein is used for *B. microti* parasites to develop.

The question then arises: what are the major energy sources for *B. microti* parasites during their intraerythrocytic development? Nutrients from the extracellular environment were thought to be the energy sources for *B. microti* parasites including glucose, and small molecules[10,35]. One possible candidate is intracellular glucose or ATP, since the amount of non-Hb proteins inside RBCs is negligible (less than 4% in wt/wt)[36]. To address this intriguing question, we measured the amount of cytoplasmic ATP in *Bm*-RBCs by measuring the bioluminescence light released from the reaction between lucifer/luciferase with ATP (Materials and Methods). The result is shown in Fig. 5. Healthy RBCs show the cytoplasmic ATP level of $2.2 \pm 0.1 \times 10^{-10}$ moles for $8 \times 10^6$ cells, which corresponds to the physiological level of cytoplasmic ATP in RBCs[37]. For *Bm*-RBCs, the measured ATP level decreased significantly to $1.5 \pm 0.2$ moles, making a 32% decrease. Considering that the parasitemia for measured *Bm*-RBCs is approximately 30%, this 32% decreased in cytoplasmic ATP suggests that nearly all of the cytoplasmic ATP is consumed in *Bm*-RBCs. Considering that cytoplasmic ATP in RBCs can be converted from D-glucose via glycolytic pathways, in the blood circulation system, metabolic energy can be continuously supplied to the *B. microti* parasite from D-glucose transported into RBC cytosol. In addition, this depletion of ATP in *Bm*-RBCs can explain the loss of the dimple shapes observed in the phase maps of *Bm*-RBCs (Figs. 3A-C).

**Mechanical alterations in *Bm*-RBCs membrane**

The amplitude of the measured dynamic membrane fluctuations of *Bm*-RBCs with PVs are comparable to those of healthy RBCs (Fig. 3G). This is also in sharp contrast to *Plasmodium* spp. in which the mechanical properties of host RBCs are significantly altered and stiffened by the production of malarial proteins, including RESA and KHARP[38-40]. In addition, a previous study has reported that *B. bovis* infected RBCs become more rigid than uninfected RBCs[41].

However, this result does not directly indicate that mechanical properties of *Bm*-RBCs are unchanged because (i) large tensional energy should be applied onto the membrane of *Bm*-RBCs due to the inclusion of large and rigid parasites, and (ii) the level of ATP which can served as a driving for membrane fluctuation should be decreased in *Bm*-RBCs. Considering the fact that the invasion of *B. microti* causes the production of large PVs inside host cells, increasing the volumes of cells and thus making the cells more sphere-like, intuitively the dynamic membrane fluctuations of sphered *Bm*-RBCs should be decreased if the mechanical properties of the *Bm*-RBCs are unchanged. This occurs because more tensional forces are applied to sphered *Bm*-RBCs compared to healthy discocytes, less fluctuation is expected similarly to the fact that sphered RBCs under hypotonic conditions exhibit less dynamic fluctuations[32]. Considering that these two factors both decreases the amplitude of dynamic membrane fluctuations, it is likely that the membrane of *Bm*-RBCs become softer and more deformable than those of healthy RBCs. This observation may open several questions about the roles of ATP in the infection of *B. microti*.

In summary, employing optical micro-tomographic technique, the structural, chemical, and mechanical alterations in *Bm*-RBCs are quantitatively and systematically investigated. The present method provides unique advantages for the study of *Bm*-RBCs, including non-invasiveness, no need for exogenous labeling agents, and quantitative imaging. Compared to healthy RBCs, *B. microti* parasitization induces an enlargement of the host cell size mainly due to the formation of PVs, the increase of the total dry mass, whereas the mass concentration in the cell cytoplasm remains unchanged. The morphology of *in situ* PVs in *Bm*-RBCs was also investigated. These characterizations were performed at the individual cell level, where it was observed that *Bm*-RBCs without PVs also undergo significant alterations, or bystander effects. In addition, dynamic membrane fluctuations of *Bm*-RBCs were simultaneous quantified with nanometer sensitivity, with indicating that no stiffening in the cell membrane occurs after a *B. microti* invasion; instead, it was suggested that the cell membrane in *Bm*-RBCs even become more deformable than those of healthy RBCs considering the decreased level of cytosolic ATP. Furthermore, our result implies that *B. microti* parasites utilize cytosolic ATP as an energy source for reproduction and development, not the Hb protein as *Plasmodium* spp. do. The present optical method for the study of *Bm*-RBCs, and the characterization of host cells, i.e., *in situ* PVs at the individual cell level, is thus expected to open new ways to unravel the pathophysiology of *Babesiosis* in various aspects as performed in the study of *Plasmodium* spp., from invasion and egress processes[42], to drug-test assay[43], to genetic destruction assay to elucidate the mechanisms of proteins that the parasite exports[38].

## Methods

**Sample preparation of mouse *Bm*-RBCs infection and determination of parasitemia** The animal care and use protocol adheres to Korean Laboratory Animal Act (Act No. 902 & 93). All experimental protocol were approved by the Committee on the Ethics of Animal Experiments of the Korean Centers for Disease Control & Prevention (permit number KCDC-035-13-2A). The use of experimental animals were maintained and handled in strict accordance with institutional guidelines of the Committee on the Ethics of Animal Experiments of the Korean Centers for Disease Control & Prevention. The procedure of *B. microti* infection in mice was complied with the institutional guidelines and regulations, with the best effort to minimize the number of animals used. *B. microti*, strain: Peabody mjr (ATCC PRA-99, USA) was maintained in Balb/c mice (KOATECH, Gyeonggi-do, Korea) before infection for this study. To passage the strain, the *B. microti* infected blood was removed from the infected mouse using cardiac puncture using a syringe and transferred rapidly in EDTA tube. For the infection, 100 μL of infected blood (20~30% parasitemia) suspension were injected intraperitoneally into Balb/c mice. Control mice were injected with an equal total number of uninfected cells. Total three healthy mice as control and three *B. microti* infected mice were used. Parasitemia after *B. microti* infection was identified by microscopic visualization of thin blood smears of tail blood stained with 5% Giemsa solution (Merck, Darmstadt, Germany). The percentage of parasitemia was determined by counting the number of infected red blood cells (RBCs) versus the total number of RBCs: % parasitemia = (infected RBCs / total RBCs) × 100. A minimum of 500 RBCs were counted and an RBC infected with multiple parasites was counted as a single infected cell. For optical measurement, control and *B. microti* infected blood were 300 times diluted in DPBS buffer. In addition, CBC blood test for healthy and infected blood were performed using automated hematology analyzer (MEK-6450, NIHON JOHDEN, Japan). MCV, MCHC and MCH were used among the parameters from CBC results.

**Common-path diffraction Optical tomography (cDOT)**

For 3-D RI tomography, optical field was measured base on the common-path laser-interferometric microscopy and optical diffraction tomography[13]. Briefly, a laser beam from a diode-pumped solid state (DPSS) laser ($\lambda$ = 532 nm, 50 mW, Cobolt, Solna, Sweden) was illuminated with various angles of illumination to a sample The sample, diluted blood sandwiched between two cover glasses, is placed between the condenser lens [UPLSAPO 60×, numerical aperture (NA) = 0.9, Olympus, Japan] and objective lens (UPLSAPO 60×, NA = 1.42, Olympus, Japan). The angle of incident beam was controlled by rotating a two-axis galvanometer mirror (GM1) (GVS012/M, Thorlabs, USA). The diffracted beam from a sample was also passing through second galvanometer (GM2), which was located at the conjugated plane of sample and synchronized with GM1 that the reflected beams have same optical path regardless of illumination angle. After GM2, the diffraction phase microscopy based on common-path interferometry was used to collect the diffraction beam from sample. Using a diffraction grating (70 grooves mm$^{-1}$, #46-067, Edmund Optics Inc., NJ, USA), the 1$^{st}$ order beam as a sample beam was interfered with spatially filtered 0$^{th}$ order beam that serves as a reference. Then, this spatially modulated interferograms were recorded on high-speed sCMOS camera (Neo sCMOS, ANDOR Inc., Northern Ireland, UK) having 528 × 512 resolution with a frame rate of 125 Hz while the incident beam was scanning spirally with 600 different angles. The total magnification of cDOT was 240 by an additional 4-$f$ system. From measured hologram images, 3-D RI distribution of sample was reconstructed via optical diffraction tomography algorithm, found elsewhere[15]. For the measurements of 2-D height profile and membrane fluctuation dynamics, the fixed normal incident angle was used to generate a hologram and it was recorded by the camera with a frame rate of 125 Hz for 2 sec. Alternately, quantitative phase imaging unit can also be used[44,45]; an existing optical microscope can be converted into a 3-D holographic microscope to investigate *Bm*-RBCs.

**Analysis of the red cell indices** The six of red cell indices are comprised of morphological (surface area, volume, and sphericity), chemical (Hb concentration and content), and mechanical (membrane fluctuation) parameters.

To measure the morphological parameters, we used the reconstructed 3-D RI maps by the diffraction optical tomography algorithm from measured multiple optical phase maps corresponding to various illumination angles on the sample. The whole volume of host *Bm*-RBCs were calculated by integrating all voxels inside individual *Bm*-RBCs. The space corresponding to cytoplasm of a RBC was selected by RI with a higher value than threshold. The threshold was defined by 50% of RI difference between the maximum RI of the cell $n_{cell\_max}$ and surrounding medium $n_m$ for determine the cell boundary, i.e. $n_{thresh} = n_m + 0.5 \cdot (n_{cell\_max} - n_m)$. For the *Bm*-RBCs with apparent PVs, voxels of inclusion PVs inside cytoplasm were also counted. Then, the total number of voxels was multiplied by the magnification of the optical system to translate in a length scale. Next, for surface area measurements, the isosurfaces of individual RBCs were reconstructed from volume data of 3-D RI maps using MATLAB. Surface area of isosurface is measured through the sum of the areas of all the patch faces, which are broken down into small triangular pieces. In addition, the sphericity index *SI*, a dimensionless quantity ranging from 0 to 1, was obtained by calculating $SI = \pi^{1/3}(6V)^{2/3}/A$ where the *V* is the volume and *A* is the surface area[14,46,47].

For measurement of the cellular dry mass, the measured 2-D phase at the normal angle was used. The total dry mass of a RBC was obtained from the integrating 2-D optical phase over entire cell area with RI increment of proteins as[47],

$$CM = \frac{\lambda}{2\pi\alpha} \int \Delta\phi(x,y) dA, \qquad (1)$$

where *CM* is the cellular dry mass, $\lambda$ is the wavelength of laser light (532 nm), $\alpha$ is RI increment (0.2 mL/g)[24-26] and $\Delta\phi(x,y)$ is 2-D optical phase. In addition, the total mass concentration in a RBC was obtained from the total dry mass divided by the cellular volume. The total dry mass quantifies all non-aqueous materials in RBC cytoplasm including Hb and non-Hb proteins.

To measure the mechanical parameter, we calculated the dynamic membrane fluctuation from the successively measured the instantaneous height map *h(x,y;t)*, given as:

$$h(x,y;t) = \frac{\lambda}{2\pi \cdot \Delta n} \Delta\phi(x,y;t), \qquad (2)$$

The values for the membrane fluctuation were calculated by averaging the root-mean-square of height displacement over the cell area and given as:

$$\Delta h(x, y) = \sqrt{\left\langle \left[ h(x, y; t) - h_m(x, y) \right]^2 \right\rangle}, \qquad (3)$$

where $h_m$ is the time averaged height at the cell surface.

**Light measurement assay for ATP amount in RBCs**

30% parasitemia of *B. microti* infected blood and non-infected blood was collected from Balb/c mouse. RBCs were isolated using Histopaque-1077 from 200 μL blood (Sigma-aldrich, Saint Louis, USA). Briefly, anti-coagulated blood is layered onto Histopaque-1077 and centrifuge at 400 *g* for precisely 30 minutes at room temperature. During centrifugation, erythrocytes are aggregated by polysucrose and speedily sediment. Each RBCs were counted to $8 \times 10^6$ cells/mL and sonicated shortly. All samples were measured using ATP bioluminescent somatic cell assay kit (Sigma-aldrich, Saint Louis, USA). 100 μL of ATP Assay Mix Working Solution was added to a reaction vial at room temperature for 3 minutes. During this period any endogenous ATP will be hydrolyzed. To a separate vial containing 100 μL of 1× Somatic Cell ATP Releasing Reagent, 50 μL of ultrapure water was added, and then 50 μL of the cell sample to be assayed was added and immediately measure the light to be emitted $I_S$. The amount of ATP released was determined by following. Both 50 μL of an appropriate ATP Standard ($0.5-10 \times 10^{-10}$ mole) and 50 μL of the cell sample to be assayed $I_{S\&I}$ were added to a 100 μL of 1× Somatic Cell ATP Releasing Reagent. For light measurement, only 100 μL samples were transferred to the reaction vial, and instantly measured the amount of light emitted with a luminometer (MicroLumat Plus LB 96V, Berthold technologies, USA). The amount of ATP released was determined by running an internal standard. The ATP concentration in each sample was calculated according to the following formula: $N_S = [N_{S\&I} \times I_S] / [I_{S\&I} - I_S]$, where $N_S$ and $N_{S\&I}$ are the number of ATP in the cell sample and a sample with an internal standard (in moles), respectively.

**ACKNOWLEDGEMENTS**

This work was supported by from National Research Foundation (NRF) of Korea (2012R1A1A1009082, 2013R1A1A3011886, 2014K1A3A1A09063027, 2012-M3C1A1-048860, 2013M3C1A3063046, 2014M3C1A3052537), Korea Centers for

Disease Control and Prevention (4847-311-210-13, 2013), KAIST-Khalifer Seed Money Project, and APCTP. K.K. is supported by Global Ph.D. Fellowship from NRF.

**AUTHOR CONTRIBUTIONS**

Y.K.P., and S.-E.L. developed the experimental idea. H.J.P performed the optical measurement & S.H.H. developed the mouse model. S.-H.C., W.-J.L. provided analysis tools and analyzed the data. K.H.K. and Y.C.K. built the optical setup and established the image processing algorithm. S.-E.L and Y.K.P. conceived and supervised the study. All authors discussed the experimental results and wrote the manuscript.

**COMPETING FINALCIAL INTERESTS**

The authors declare no competing financial interests.

Figures with captions

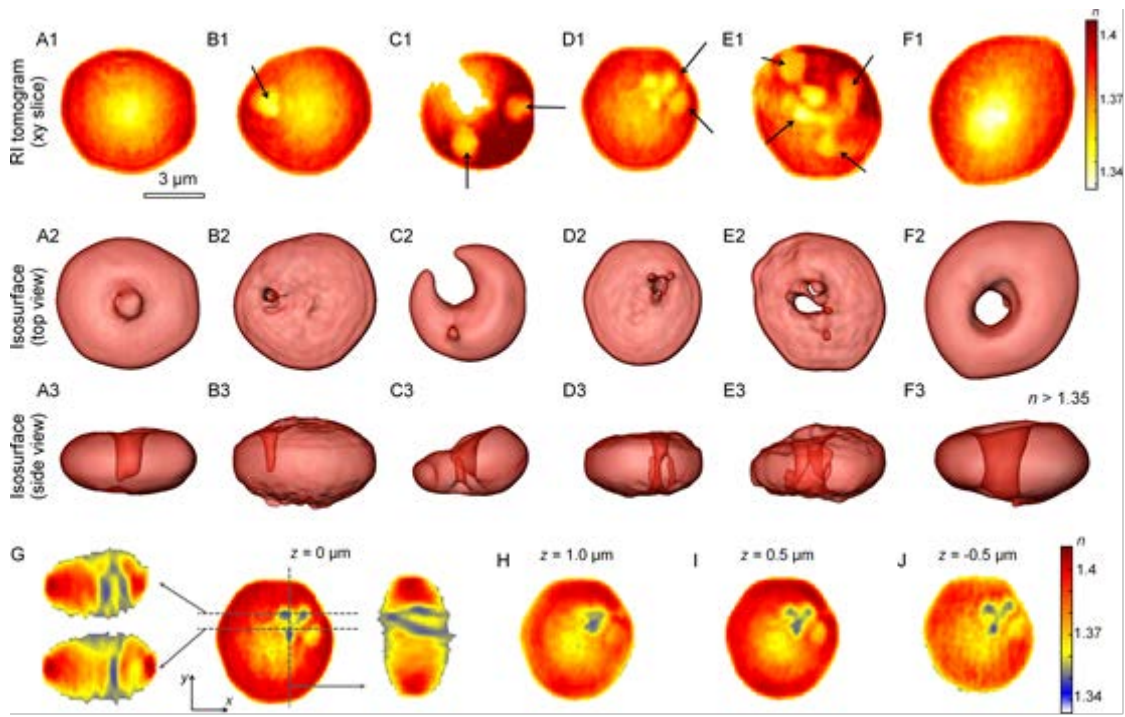

**Fig. 1 3-D RI maps reveals organization of *B. microti* parasites inside individual RBCs.** (A1-F1) The cross-sections of 3-D RI maps of individual RBCs at focus plane: (left to right) a healthy RBC, *Bm-RBCs* with single, double, quadruple, multiple infection sites and without parasite vacuoles, respectively. (A2-F2 & A3-F3) 3-D rendered isosurface of RI maps from the top and side view, respectively. (G-J) The cross-sections of 3-D RI maps of a *Bm*-RBC with quadruple infection sites (D1) at the focal plane ($z = 0$, *x-y* focus plane), *y-z* plane at 0.87 μm (right side), *x-z* plane at $y = 0.5$ μm (left bottom) and $y = 1.4$ μm (right top). (H) The *x-y* plane at 1.0 μm above the focus. (I) The *x-y* plane at 0.5 μm above the focus. (j) The *x-y* plane at 0.5 μm below the focus. The color bars are for RI. The scale bar is 3 μm.

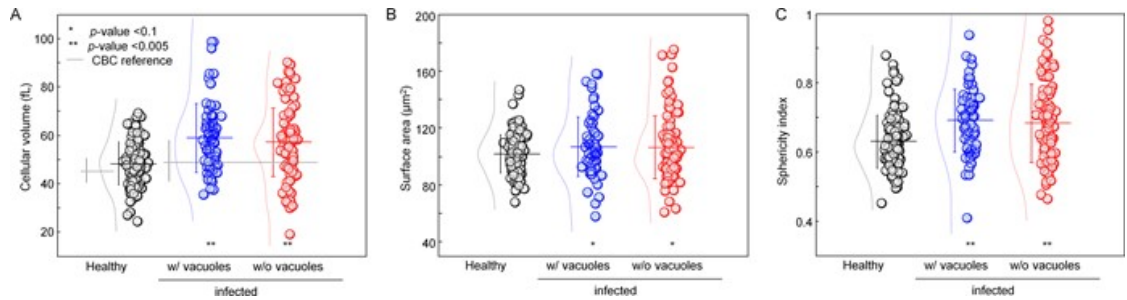

**Fig. 2** Red cell indices for healthy RBCs ($N = 123$), *Bm*-RBCs with occupied parasite vacuoles ($N = 69$) and without parasite vacuoles ($N = 102$). (A) Cellular volume, (B) cellular surface area, and (C) sphericity index of RBCs. Each symbols represent individual RBC measurements and the horizontal lines are mean values with vertical lines of standard deviation error bars. Histograms of each index with a Gaussian distribution fit are also represented next to symbols. Gray dotted lines in (A) correspond to MCV from the CBC blood test with RDW for error bar.

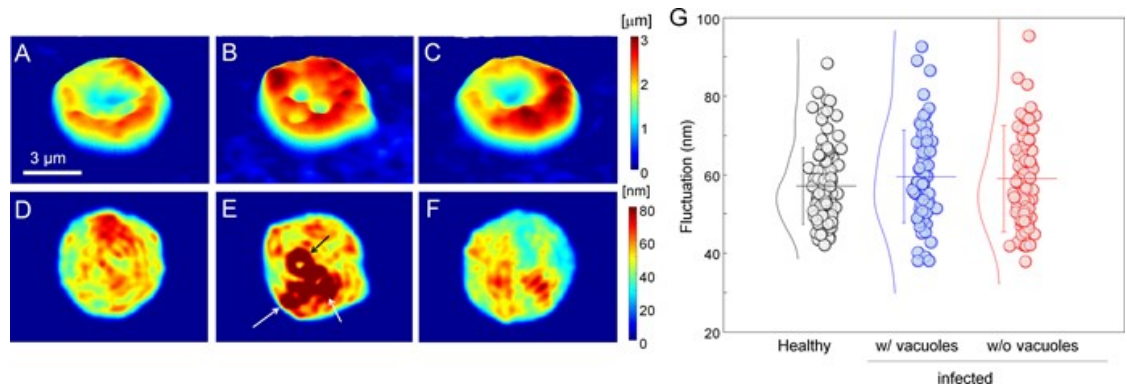

**Fig. 3 Topographic images and cell elasticity maps of *Bm-RBCs*.** **(A-C)** 2-D height profiles of a healthy RBC (A) and a *Bm*-RBC with occupied parasite vacuoles (B) and without parasite vacuoles (C). Dynamic fluctuations in the cell membrane of a healthy RBC (D) and a *Bm*-RBC with occupied parasite vacuoles (E) and without parasite vacuoles (F). (G) Membrane fluctuation of healthy RBCs, *Bm*-RBCs with occupied parasite vacuoles, and without parasite vacuoles.

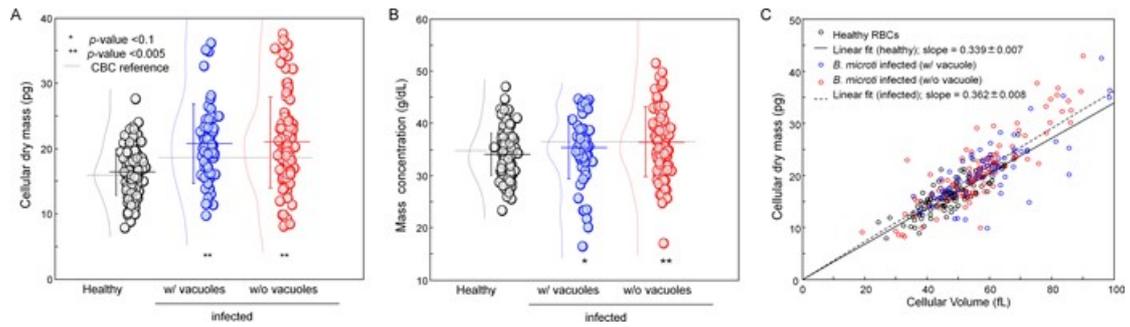

**Fig. 4 Cellular dry mass and mass concentration of *Bm*-RBCs.** (A-B) cellular dry mass and mass concentration measured for healthy RBCs ($N = 123$), *Bm*-RBCs with occupied parasite vacuoles ($N = 69$) and without parasite vacuoles ($N = 102$). Each symbols represent individual RBC measurements and the horizontal lines are mean values with vertical lines of standard deviation error bars. Histograms of each index with a Gaussian distribution fit are also represented next to symbols. Horizontal gray dotted lines in (A, B) correspond to MCH and MCHC from the CBC blood test, respectively. (C) Correlation map of cellular volume and cellular dry mass.

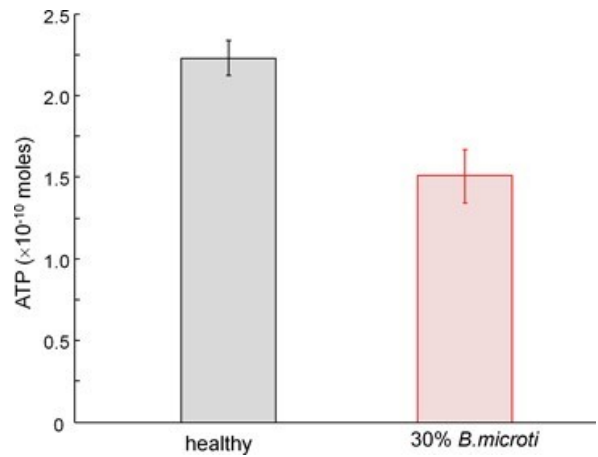

**Fig. 5 Consumption of ATP in *B.microti* infection RBCs.** Experimental measurements of the photon emission rate resulting from the reaction between luciferase/luciferin and ATP, measured for healthy and *Bm*-RBCs with 30% parasitemia.